\documentclass{svjour3}          
\usepackage{graphicx}
\usepackage{adjustbox}
\usepackage{amsmath}
\usepackage[graphicx]{realboxes}
\usepackage{rotating}
\usepackage{epstopdf}
\usepackage{float}
\usepackage{latexsym}
\usepackage{amssymb}
\usepackage{array}
\usepackage{longtable}
\usepackage{dcolumn}
\usepackage{bm}

\begin{document}

\title{On the possibility of  traversable wormhole formation in the Galactic halo in the presence of scalar field}

\author{B. C. Paul }

\institute{	
             B. C. Paul \\
             \email{bcpaul@associates.iucaa.in}\at
              Department of Physics, University of North Bengal, Siliguri, Dist. : Darjeeling 734 013, West Bengal, India \\
             }

\date{Received: date / Accepted: date}

\maketitle

\begin{abstract}
In the paper we obtain   traversable wormhole (TW) solutions in the Einstein gravity with a functional form of the   dark matter in the galactic halo. The dark matter model is  pseudo isothermal which is derived from modified gravity.  For a given central density the possibility of TW is explored determining the shape functions.  The null energy condition (NEC) and the weak energy conditions in the model is probed. We also study the existence of TWs considering homogeneous  scalar field in addition to the dark matter halo in the galaxies. An interesting observation is that  TW solutions  exist even if  NEC is not violated in the presence of scalar field.

\end{abstract}

\maketitle	

\section{ Introduction}
\label{intro}
General Theory of Relativity (GTR) permits wormhole solutions of the Einstein field equations for a special composition of matter different from perfect fluid. Wormholes are tunnel like objects which may connect different spacetime regions within the universe, or different universes \cite{1,2}.
In recent times the study of
traversable wormholes  (TW)  has attracted a lot of interest with the pioneering work of Ellis \cite{7a,7b}, Bronnikov \cite{7c} including  the seminal work of Morris and Thorne \cite{7d}.
 The geometry of TW requires exotic matter concentrated at the wormhole throat implying violation of   the energy conditions,  namely  null
energy condition (NEC) \cite{12}. It is speculated that the exotic matter can exist in the context of quantum field theory.
The concept of dark matter, a hypothetical form of matter that makes up about 25 \% of the matter composition of the Universe is predicted from observations. A number of possible candidates from particle physics and supersymmetric string theory such as axions
and weakly interacting massive particles (WIMP) are considered  to describe as the candidate of the dark matter though there is  no direct experimental detection of dark matter. Nevertheless, its existence are hinted in 
the galactic rotation curves \cite{13}, the galaxy clusters dynamics  \cite{14},  and also at cosmological scales
of anisotropies encoded in the cosmic microwave background measured by PLANCK  \cite{15}.

In the literature \cite{r14,r15,r16} considering characteristics of dark matter halos and galaxies  formation of traversable wormholes are considered based on the Navarro-Frenk-White (NFW) profile \cite{nfw} of matter distribution. 
TWs may form in the outer halo of galaxies which are interesting to describe time machines.
Rahaman et al. \cite{31} first proposed the possible existence of wormholes in the outer regions of the galactic
halo based on the NFW density profile, then they used the Universal Rotation Curve (URC)
dark matter model to obtain analogous results for the central parts of the halo \cite{32}, DM has been adapted as a non-relativistic phenomenon such as NFW profile and King profile to study WH construction \cite{33}.
Recently, Jusufi {\it et. al.} \cite{r17} pointed out the possibility of traversable wormhole formation with a Bose-Einstein condensation dark matter halo, however, in this case approximate solutions are found.
The existence of TW in the spherical stellar systems based on the fact that the
dark matter halos produced in computer simulations are
best described by such a profile is also investigated. Assuming  a number of parameters that exists in the Einasto profile \cite{ei1,ei2} it is shown that TWs in the outer regions of spiral galaxies are possible while the inner part regions prohibits such formations.  The Bose-Einstein Condensation dark matter (BEC-DM) model shows that the model is more realistic on the small scales of galaxies compared to the CDM model. For instance, the interactions
between dark matter particles are very strong in the inner regions of galaxies, and thus the dark matter will no longer
be cold. For the BEC-DM model, the dark matter density profile can be described by the Thomas-Fermi (TF) profile  \cite {cg}. The  possibility of  formation of TWs in the dark matter halo is  studied with isotropic pressure with TF profile \cite{cg1}.
The BEC-DM model predicts much less dark matter density in the central regions of galaxies compared to that found with the NFW profile. There exists another class of DM described by pseudo isothermal (PI) profile in addition to the CDM model and the BEC-DM model which is associated with the modified gravity, such as Modified Newtonian Dynamics (MOND) \cite{mond}.  In the MOND model, the dark matter density
profile is given by
\begin{equation}
\label{e1}
\rho_{DM} = \frac{\rho_0}{1+ \left( \frac{r}{R_c} \right)^2}
\end{equation}
where $\rho_0$ is the central dark matter density and $R_c$ is the scale radius.
Using the above formalism, we adapted the DM density distribution and used to study wormhole construction for MOND. Subsequently we introduce homogeneous scalar field in addition to MOND dark matter to investigate the  TW models and the matter required for their existence.

The present paper is organized as follows. In Section II, the Einstein gravity and field equations are obtained. The traversable wormhole solutions are obtained with MOND dark energy profile. In Section III, the existence of wormhole wth MOND and scalar field is discussed and finally a brief discussion in Section IV.

\section{Einstein Field Equation and Traversable Wormhole solution}

The Einstein field equation is given by 
\begin{equation}
\label{e2}
R_{\mu \nu} -\frac{1}{2} g_{\mu \nu} R = T_{\mu \nu}
\end{equation}
where $\mu, \nu = 0,1, 2,3$ and  $T_{\mu \nu }= diag ( -\rho, P_r,  P_t, P_t)$ is the energy-momentum tensor, in gravitational unit  $ c=8 \pi G=1$.  The spacetime metric of a spherically symmetric traversable wormhole  is given by
\begin{equation}
\label{e3}
ds^2= -e^{2 \Lambda (r)} dt^2 + \frac{dr^2}{1- \alpha(r)} + r^2 (d \theta^2 + sin^2 \theta d \phi^2)
\end{equation}
where $\Lambda $ is the redshift function.  For wormhole we consider  
$\alpha (r)= \frac{b(r)}{r}$, where $b(r)$ is the shape function. In order to ensure a wormhole to be traversable, there should be no event horizon.  In this case  $\Lambda (r)$  should be finite which tends to zero when $ r \rightarrow \infty$. The geometry of the
wormhole is determined by the shape function $b(r)$, which at the throat of the wormhole $r=r_0$ should satisfy the condition $b(r_0) = r_0$. To keep the wormhole's  throat open, the shape function $b(r)$ should satisfy the flare out condition $\frac{b(r) - r b'(r)}{b^2(r)} > 0$  and $b'(r) < 1$.

Using the Einstein field eqs. (\ref{e2}) and  the metric given by eq. (\ref{e3}), we get
\begin{equation}
\label{e4}
\rho (r)= \frac{b'(r)}{8 \pi r^2} 
\end{equation}
\begin{equation} 
\label{e5}
 8 \pi P_r (r) = - \; \frac{b(r)}{r^3} + 2 \left( 1- \frac{b(r)}{r} \right) \frac{\Lambda'(r)}{r}       
 \end{equation}
\begin{equation}
\label{e6}
8 \pi P_t (r)  = \left(1- \frac{b}{r} \right)  \left[ \Lambda''(r) +  \Lambda'^2 
 - U  \Lambda' -  \frac{U}{r}   + \frac{   \Lambda' }{r}\right]
\end{equation}
setting $U= \frac{ r b'(r) -b(r)}{2 r(r-b(r))}$ where  $\rho(r)$ is the energy density for dark matter, $P_r$ and $P_t$ are radial and transverse pressures. 
For a flat rotation curve for the circular stable geodesic motion in the equatorial plane of a galaxy one finds
\begin{equation}
\label{e7}
e^{2 \Lambda (r)} = E_0 \; r^l.
\end{equation}
In the above $l= 2 (v^{\phi})^2$ where $v^{\phi}$ represents the rotational velocity and $E_0$ is an integration constant for a large $r$ \cite{32,kn,ca}.  For a typical galaxy it is found that $v^{\phi} \sim 10^{-3}$ (300 km/s)  within 300 kpc \cite{nca}. 
 We assume that the density of the wormhole matter is described  by the density profile of the MOND. We consider 
critical value $E_0= \frac{1}{R_c^l}$ where $R_c$ is the scale radius, the spacetime metric can be written as
\begin{equation}
\label{e8}
ds^2= - \left(\frac{r}{R_c}\right)^l dt^2 + \frac{dr^2}{1- \frac{b(r)}{r}} + r^2 (d \theta^2 + sin^2 \theta d \phi^2)
\end{equation}
Solving Einstein field eq. (\ref{e4})  using the DM density given  by eq. (\ref{e1}), we obtain
\begin{equation}
\label{e7}
b(r) = \rho_0 R_c^3 \left[ \frac{r}{R_c} - tan^{-1} \frac{r}{R_c} \right] +C
\end{equation}
where $C$ is an integration constant. 

\begin{figure} 
\includegraphics[scale=0.6]{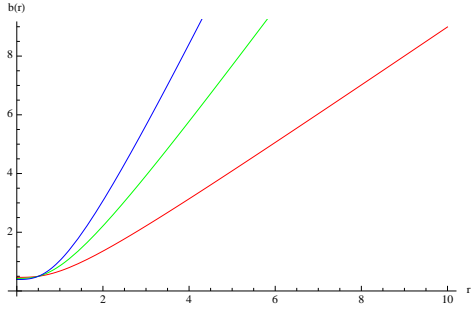}
\caption{Variation  $b(r)$ with $r$ in the unit of $R_c$  for $\rho_0 R_c^3=1$ (red), $\; 2 $ (green),  $3 \; (blue)$ }
\label{Fig: 1}
\end{figure}
\begin{figure} 
\includegraphics[scale=0.6]{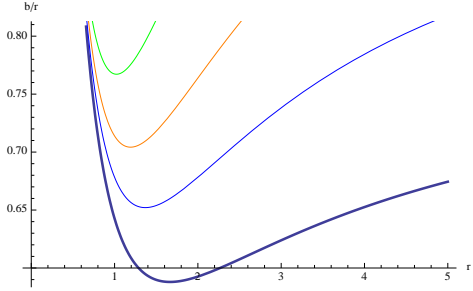}
\caption{Radial variation  of $\frac{b(r)}{r}$  for $\rho_0 \;R_c^3 ==0.8$ $(thick line)$,  $1 \;(blue)$, $1.2 \; (orange)$, $1.5\; (green)$ }
\label{Fig: 2}
\end{figure}
\begin{figure} 
\includegraphics[scale=0.6]{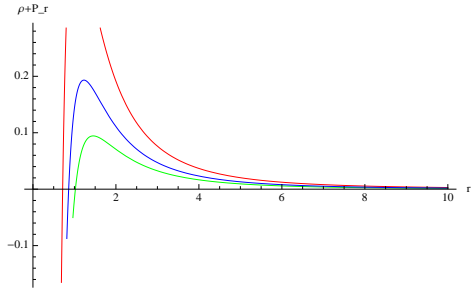}
\caption{Variation  NEC with $r$ in the unit of  $R_c$  with $\rho_0  R_c^3=1$ for $l= 0.5 \; (green)$ \, $1\; (blue)$, $2 \;(red) $}
\label{Fig: 3}
\end{figure}

Using the boundary condition $b(r_0)=r_0$, the integration constant $C$ can be fixed for the shape function which is given by
\begin{equation}
\label{e8}
b(r) = \rho_0 R_c^3 \left[ \frac{r}{R_c} -  \frac{r_0}{R_c} - \left(tan^{-1} \frac{r}{R_c} - tan^{-1} \frac{r_0}{R_c} \right)\right] +r_0,
\end{equation}
where $r_0$ represents the throat of the wormhole.
The flaring out condition is required to be valid to keep the mouth of the wormhole open which is satisfied here as
\begin{equation}
\label{e9}
b'(r) = \rho_0 R_c^2 \; \frac{ \frac{r^2}{R_c^2}}{\left(1+\frac{r^2}{R_c^2} \right)} <1
\end{equation}
for $\rho_0 R_c^2 \sim 1$. In Fig. (1) we plot radial variation of  $b(r)$ with different $ \rho_0 \;R_c^3$. It is found that as the central density is increased $b(r)$ is found to  increases for a given radial distance.
 The radial variation of the flaring out condition is checked in Fig. (2) it is found that $\frac{b(r)}{r} <1 $ as long as $\rho_0 R_c^3 \leq 1$.
 
 Next let us discuss the issue of energy conditions and make some regional plots to check the validity of the energy
conditions. Recall that the WEC \cite{12}  is defined by $T_{\mu \nu} U^{\mu} U^{\nu} \geq 0$ $i.e., $
\begin{equation}
\label{e10}
\rho \geq 0,  \; \; \;  \rho(r) +P_r \geq 0
\end{equation}
where $U^{\mu}$ denotes the time like vector. This means that local energy density is positive and it gives rise to the continuity of NEC, which is defined by $ T_{\mu \nu} k^{\mu} k^{\nu} \geq 0$ $i.e., $
\begin{equation}
\label{e11}
 \rho(r) +P_r \geq 0
\end{equation}
where $k^{\mu}$ represents a null vector.
  The NEC is checked in Fig. (3), it is found that near the throat NEC is not obeyed but away from the throat it always satisfied which is shown for different power law index ($l$) in eq. (7).  We note that for $l=2$, NEC is always obeyed for the central density $\rho_0  < \frac{1.2}{R_c^3}$. It is evident from Fig. (4) that traversable wormholes exists with exotic matter near the throat but away from the throat  exotic matter is not required  in a low central density galaxy
 but for higher central density galaxies initially there is exotic matter then again exotic matter followed by normal matter at $r \rightarrow \infty$.

\begin{figure} 
\includegraphics[scale=0.6]{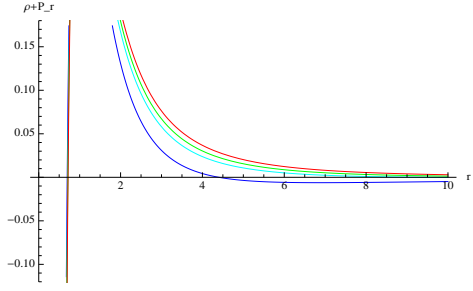}
\caption{Variation  NEC with $r$ for $R_c  $  with $l=2$ for $\rho_0 \;R_c^3 = 1 \;  (red)$, $1.1 \; (green) $,  $1.2 \;(cyan) $, $1.5  \; (blue)$ }
\label{Fig: 4}
\end{figure}

\begin{figure} 
\includegraphics[scale=0.6]{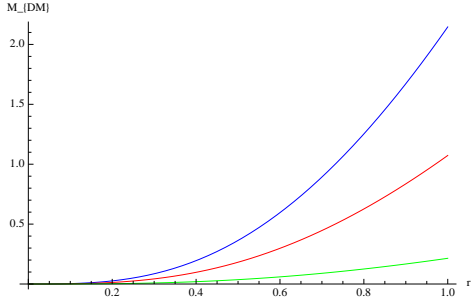}
\caption{Variation  mass profile with $r$ in unit of $R_c$  with $4 \pi \rho_0 R_c^3=1 \;(green)$,  $5 \;  (red)$, $10 \; (Blue) $ }
\label{Fig: 5}
\end{figure}
The mass profile of the galactic halo is
\begin{equation}
M_{DM} (r)= 4 \pi \int_0^r \rho_{DM} (r') r'^2 dr'
\end{equation}
which yields
\begin{equation}
\label{18}
M_{DM} (r)= 4 \pi \rho_0 R_c^3 \left[\frac{r}{R_c} - tan^{-1} \left(\frac{r}{R_c} \right) \right].
\end{equation}
In Fig. (5), it is observed that  dark matter increases with the increase in radial size in unit of $R_c$ for increasing values of  $4 \pi \rho_0 R_c^3 = m_0$. It is evident that as the mass parameter ($m_0$) is increased 
From the eq. (\ref{18}),  one can find the tangential velocity \cite{har} $v_{tg}^2 (r) = \frac{GM_{DM}(r)}{r}$ for a test particle moving in the  dark halo given by 
\begin{equation}
\label{19}
v_{tg}^2 (r)= 4 \pi \rho_0 R_c^{2} \left[1 - \frac{R_c}{r} \; tan^{-1} \left(\frac{r}{R_c} \right) \right],
\end{equation}
which represents a constant rotational curve away from the centre of the galaxy.
The rotational velocity of a test particle within the equatorial plane is determined by 
\begin{equation}
\label{20}
v_{tg}^2 (r) = r \Lambda' (r)
\end{equation}
Combining eq.(\ref{20}) with the expression for the test particle moving in the dark halo given by eq. (\ref{19}) we get
\begin{equation}
 r \Lambda' (r) = 4 \pi \rho_0 R_c^{2} \left[1 - \frac{R_c}{r} \; tan^{-1} \left(\frac{r}{R_c} \right) \right].
\end{equation}
On integrating 
\begin{equation}
\Lambda (r) = 4 \pi \rho_0 R_c^{2} \left[ ln \frac{ \frac{r}{R_c} }{ \sqrt{1+ \left( \frac{r}{R_c} \right)^2}}
  -  R_c \; \frac{tan^{-1} \left(\frac{r}{R_c} \right)}{r}  \right] +C_1
\end{equation}
where $C_1$ is an integration constant. Thus the $g_{00}$ component of the metric tensor become
\begin{equation}
e^{2\Lambda (r) } = D e^{ 8 \pi \rho_0 R_c^{2} \left[ ln \frac{ \frac{r}{R_c} }{ \sqrt{1+ \left( \frac{r}{R_c} \right)^2}}
  -  R_c \; \frac{tan^{-1} \left(\frac{r}{R_c} \right)}{r}  \right] + 2 C_1},
\end{equation}
the new integration constant $D$ can be absorbed by rescaling $t\rightarrow Dt$. For finite redshift we get a constraint equation which determines $C_1$ as follows
\[
lim_{r \rightarrow R_c} e^{2\Lambda (r) } =1.
\]
The limiting value $$lim_{ r \rightarrow R_c}   e^{ 8 \pi \rho_0 R_c^{2} \left[ ln \frac{ \frac{r}{R_c} }{ \sqrt{1+ \left( \frac{r}{R_c} \right)^2}} -  R_c \; \frac{tan^{-1} \left(\frac{r}{R_c} \right)}{r}  \right] + 2 C_1} =1$$ determines the constant as
\begin{equation}
C_1= 2 \pi R_c^2\left( \frac{\pi}{2}+ln \; 2\right).
\end{equation}

The wormhole metric is not asymptotically flat, therefore, we  employ matching conditions by truncating the wormhole metric at radius $R$ and connecting with the exterior Schwarzschild black hole metric  as the later corresponds to vacuum solution and asymptotically flat. Now imposing the matching condition at $r=R$, we get
\begin{equation}
\label{e20}
e^{ 8 \pi \rho_0 R_c^{2} \left[ ln \frac{ \frac{R}{R_c} }{ \sqrt{1+ \left( \frac{R}{R_c} \right)^2}}
  -  R_c \; \frac{tan^{-1} \left(\frac{R}{R_c} \right)}{R}  \right] + 4 \pi R_c^2\left( \frac{\pi}{2}+ln \; 2\right)}= 1- \frac{2M}{R}
\end{equation}
\begin{equation}
1- \frac{b(R)}{R} = 1-\frac{2M}{R}
\end{equation}
where the active mass of the galaxy ($M$) is determined as
\begin{equation}
\label{e21}
M = \frac{  \rho_0 R_c^3}{2}  \left[ \frac{R}{R_c} -  \frac{r_0}{R_c} - tan^{-1} \frac{R}{R_c}+ tan^{-1} \frac{r_0}{R_c} \right] +\frac{r_0}{2}
\end{equation}
in the gravitational unit $\rho_0 R_c^2 \sim 1$. 
The eqs. (\ref{e20}) and (\ref{e21}) implicitly provide the value of truncated radius ($R$) where the matching occurs. Note that   $\rho_0$ represents  galactic dark matter  in the unit of $M_{DM}$ $halo/kpc^3$,  $r$ has units of $M_{DM}$ halo and $R$ and $r_0$ are in the units of $kpc$. From eq. (\ref{e20}), $R_c$ is determined for density at the throat of the wormhole knowing the central dark matter density $\rho_0$.

\section{Existence of Wormhole with MOND and scalar field}

In this section we consider wormhole solutions in the presence of scalar field and MOND dark matter density profile.
Therefore the energy momentum tensor consists of two terms $T_{\mu \nu} = T_{DM}^{\mu \nu} + T^{f}_{\mu \nu} $. To determine the second part of stress energy tensor we consider a minimally coupled massless scalar field described by the Lagrangian:
\begin{equation}
\label{e22}
\L= \frac{1}{2} \sqrt{-g} g^{\mu \nu} \phi_{;\mu} \phi_{;\nu}.
\end{equation}
The equation of motion for $\phi$ is
\begin{equation}
\label{e23}
\square  \phi =0.
\end{equation}
The stress energy tensor for the scalar field is obtained from eq. (\ref{e22}) which is
\begin{equation}
\label{e24}
T^f_{\mu \nu} = \phi_{;\mu} \phi_{;\nu} - \frac{1}{2} g_{\mu \nu} g^{\sigma \eta} \phi_{;\sigma} \phi_{;\eta}.
\end{equation}
The conservation eq. (\ref{e23}) of $\phi$ is given by
\begin{equation}
\label{e25}
\frac{ \phi "}{\phi'} + \frac{1}{2} \frac{(1-\frac{b}{r})'}{(1-\frac{b}{r})} + \frac{2}{r}=0.
\end{equation}
On integrating we get
\begin{equation}
\label{e26}
\phi'^{2} r^4 \left( 1-\frac{b}{r} \right)= \frac{\phi_0}{4 \pi}
\end{equation}
where $\phi_o > 0$. Now the Einstein  field equations are given by
\begin{equation}
\label{e31}
\frac{b'(r)}{8 \pi r^2}=  \rho (r) + \frac{1}{2} \phi'^{2} \left( 1- \frac{b(r)}{r} \right)
\end{equation}
\begin{equation} 
\label{e32}
 - \; \frac{b(r)}{8 \pi r^3} +  \left( 1- \frac{b(r)}{r} \right) \frac{\Lambda'(r)}{4 \pi r}   = P_r (r) - \frac{1}{2} \phi'^{2} \left( 1- \frac{b(r)}{r} \right)     
 \end{equation}
\[
 \frac{1}{8 \pi} \left(1- \frac{b}{r} \right)  \left[ \Lambda''(r) +  \Lambda'^2 
 - U  \Lambda' -  \frac{U}{r}   + \frac{   \Lambda' }{r}\right]
\]
\begin{equation}
= P_t (r) - \frac{1}{2} \phi'^{2} \left( 1- \frac{b(r)}{r} \right) 
\end{equation}
setting $U= \frac{ r b'(r) -b(r)}{2 r(r-b(r))}$ where  $\rho(r)$ is the energy density for dark matter, $P_r$ and $P_t$ are radial and transverse pressures. Using the solution given by eq. (\ref{e26}) in the field eq. (\ref{e31}) we get
\begin{equation}
\label{e28}
\frac{b'(r)}{8 \pi r^2}=  \rho (r) + \frac{\phi_0}{8 \pi r^4}.
\end{equation}

\begin{table}
\begin{center}
\tabcolsep=0.11cm
\begin{tabular}{|c|c|c|} \hline
$\phi_0$ & $ r$     & NEC ($r \geq r_0$)\\ \hline
0.05  & 0.376 & $\surd$ \\ \hline
0.08  & 0.406 & $\surd$  \\ \hline
0.10  & 0.422 &  $\surd$ \\ \hline
0.20  & 0.479 & $\surd$  \\ \hline
0.25  & 0.50 & 	X \\ \hline
0.26  & 0.504 & X \\ \hline
0.30  & 0.518 & X \\ \hline
\end{tabular}
\caption{NEC for $\rho_0 R_c^3 =10$ and throat radius $r_0= 0.5$, "X" means violation and "$\surd$" means satisfied}
\end{center}
\end{table}

\begin{figure} 
\includegraphics[scale=0.8]{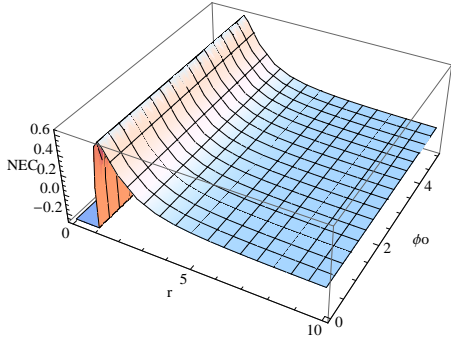}
\caption{Variation of NEC with $r$ in unit of $R_c$  with $4 \pi R_c^3=1 $ and different $\phi_0$ }
\label{Fig: 6}
\end{figure}


Integrating we determine the shape function as
\[
b(r) = \rho_0 R_c^3 \left[ \frac{r}{R_c} -  \frac{r_0}{R_c} - \left(tan^{-1} \frac{r}{R_c} - tan^{-1} \frac{r_0}{R_c} \right)\right]
\]
\begin{equation}
\label{e29}
 \; \; \; \; + \phi_0 \left(\frac{r_0-r}{rr_0} \right)+r_0 
\end{equation}
For $\Lambda \rightarrow 0$, no horizon occurs and the radial variation of  NEC is drawn in Fig. (6) for a range of values of $\phi_0$. It is found that near the throat NEC is  violated  but away from the throat it is obeyed which depends on $\phi_0$. The NEC is checked for a given $\phi_0 R_c^3$ and throat radius numerically which are tabulated in the Tables I-II. It is evident that as  $\phi_0 R_c^3$  is increased NEC is satisfied at lower $\phi_0$ of the scalar field. In Table-III, we display the values of $\phi_0$ for a given throat radius $r_0$ with density parameter of the galaxy. It is found that if the throat radius is big then  one needs a scalar field with high $\phi_0$ scalar field value for accommodating TW with normal matter for a given density of the galactic halo. It is also noted that for same throat radius if the density is high then one requires large value of the scalar field.

\begin{table}
\begin{center}
\tabcolsep=0.11cm
\begin{tabular}{|c|c|c|} \hline
$\phi_0$ & $ r$     & NEC  ($r \geq r_0$) \\ \hline
0.10 & 0.339 & $\surd$ \\ \hline
0.20  & 0.400 &  $\surd$  \\ \hline
0.30 & 0.444&  $\surd$  \\ \hline
0.50  & 0.5 &  $\surd$  \\ \hline
0.60  & 0.520 & X \\ \hline
\end{tabular}
\caption{NEC for $\rho_0 R_c^3 =8$ and throat radius $r_0= 0.5$, "X" means violation and "$\surd$" means satisfied}
\end{center}
\end{table}

\begin{table}
\begin{center}
\tabcolsep=0.11cm
\begin{tabular}{|c|c|c|} \hline
$\rho_0R_c^3$ & $r \geq$ & $\phi_0$   \\ \hline
10  & 0.5 & 0.25 \\ \hline
10 & 0.4 & 0.065 \\ \hline
8 & 0.5&0.5 \\ \hline
6 & 0.5 & 0.051 \\ \hline
5  & 0.5 & $\sim 0$ \\ \hline
\end{tabular}
\caption{Tabulation of $\phi_0$ with $\rho_0 R_c^3$ and radial distance $r$ from which NEC is satisfied.}
\end{center}
\end{table}

\section{Discussion}

In the present paper we study the possibility of traversable wormhole in the Einstein gravity with MOND dark matter profile in the presence and absence of scalar field. It is found that in the absence of scalar field one requires exotic matter at the throat of the wormhole. It is also noted that there exists two scenario (i) exotic matter at the throat and then normal matter away from the throat as NEC is obeyed and (ii) exotic matter at the throat, subsequently away from the throat NEC violates followed by the region where NEC is obeyed depending on the value of the central density. It is also found that if the central density is large the scale radius will be low for a given galactic halo mass.
In the presence of homogeneous scalar field in addition to dark matter represented by MOND, we obtain a class of TW with exotic or  without  exotic matter depending on the values of the scalar field $\phi_0$ and density of the galactic halo. The later result represents a new  class of TW not shown earlier. The TW in higher dimensional gravity \cite{pkc} is important to probe in this context which will be taken up elsewhere.

\section{Acknowledgment}
The author would like to thank IUCAA , Pune and IUCAA Centre for Astronomy Research and Development (ICARD), NBU for extending research facilities and  DST-SERB Govt. of India (File No.:EMR/2016/005734).\\

\end{document}